\documentstyle[emulateapj,psfig]{article}
\def\gs{\mathrel{\raise0.35ex\hbox{$\scriptstyle >$}\kern-0.6em
\lower0.40ex\hbox{{$\scriptstyle \sim$}}}}
\def\ls{\mathrel{\raise0.35ex\hbox{$\scriptstyle <$}\kern-0.6em
\lower0.40ex\hbox{{$\scriptstyle \sim$}}}}
\slugcomment{Accepted for publication in The Astrophysical Journal}

\lefthead{Brainerd et al.}
\righthead{A Comparison of Simple Mass Estimators for Galaxy Clusters}

\begin{document}

\title{A Comparison of Simple Mass Estimators for Galaxy Clusters}

\author{Tereasa G.\ Brainerd$^1$, Candace Oaxaca Wright$^1$, }
\author{
David M.\ Goldberg$^2$, \&
Jens Verner Villumsen$^3$}

\affil{\tiny 1) Boston University, Department of Astronomy, Boston, MA 02215}
\affil{\tiny 2) Princeton University, 
Department of Astrophysical Science, Princeton,
NJ, 08544}
\affil{\tiny 3) Current postal address: 
5151 Reed Road, Suite 224C, Columbus, OH 43220}

\begin{abstract}
High-resolution N-body simulations are used to
investigate systematic trends in the mass profiles and total masses
of clusters as derived from
3 simple estimators: (1) the weak gravitational lensing shear
field under the assumption of an isothermal
cluster potential, (2) the dynamical mass obtained from the measured
velocity dispersion under the assumption of an isothermal
cluster potential, and (3) the classical virial estimator.
The clusters used for the analysis 
were obtained from simulations of a standard cold dark matter
universe at $z=0.5$ and consist of order
$2\times 10^5$ to $3\times 10^5$ particles of 
mass $m_p \simeq 10^{10} M_\odot$.
The clusters are not smooth and spherically symmetric but, rather, 
the mass distribution is triaxial and significant
substructure exists within the clusters.  Not surprisingly,
the level of agreement between the cluster mass profiles obtained from the
various estimators and the actual mass profiles is found to be
scale-dependent.

We define the total cluster mass to be the mass contained
within a 3-dimensional radius $r_{200}$ of the cluster centers, where 
$r_{200}$ is the radius inside which the mean interior overdensity is equal to
200.  Under this definition the classical virial estimator yields a
good measurement of the total cluster mass, though it is systematically
underestimated by $\sim 10$\%.  This result suggests that, at least in
the limit of ideal data, the
virial estimator is quite robust to deviations from pure spherical
symmetry and the presence of substructure within a cluster.
The dynamical mass estimate based upon
a measurement of the cluster velocity dispersion and an assumption of
an isothermal potential yields a poor measurement of the total cluster
mass, underestimating it by $\sim 40$\% for the case that $\sigma_v$
is computed from an average over the entire cluster.  The weak lensing
estimate yields a very good measurement of 
the total cluster mass, provided the mean shear used to
determine the equivalent cluster velocity dispersion is computed from an
average of the lensing signal
over the entire cluster (i.e. the mean shear is computed
interior to a projected
cluster radius of $R_{200}$).

\end{abstract}

\keywords{cosmology: theory --- dark matter -- gravitational lensing ---
large-scale structure of the Universe --- galaxies: clusters: general ---
methods: numerical}

%%%%%%%%%%%%%%%%%%%%%%%%%%%%%%%%%%%%%%%%%%%%%%%%%%%%%%%%%%%%%%%%%%%%%

\section{Introduction}

Rich clusters of galaxies constitute the largest gravitationally-bound
objects in the universe and the history of their formation is
a potentially powerful test of the viability of differing
large-scale structure models.  In particular, the time evolution of
the cluster mass function is expected to be a good discriminator
between low- and high-density universes, and, additionally, it provides
a constraint on the degree of bias between the galaxy and mass
distribution (e.g.\ Bahcall et al., 1997; Fan et al., 1997).
Owing to
the lack of complete, uniform, mass-selected catalogs of galaxy clusters
out to large depths ($z\sim 0.5$ to $1.0$), the evolution of the cluster
mass function is not constrained especially well at present.
It is expected, however,
that significant effort will soon be devoted to developing such
catalogs, particularly using deep wide-field imaging of weak gravitational
lensing of distant field galaxies by intervening clusters.
The ability of these future catalogs to constrain models of
structure formation rests heavily on the accuracy with which the
cluster masses can be obtained and an understanding of any systematic
biases present in the mass estimators themselves.

The virial mass estimator is the method which has the longest 
history of application to galaxy clusters  
and it yields consistent results for cluster
mass to light ratios in the range of
$200h (M/L)_\odot$ to $400h (M/L)_\odot$ (e.g.\ Zwicky 1933, 1937; Smith 1936;
Schwarzschild 1954; Gott \& Turner 1976; Gunn 1978;
Ramella, Geller \& Huchra 1989; David, Jones \& Forman 1995; Bahcall,
Lubin \& Dorman 1995; Carlberg et al.\ 1996; Carlberg, Yee \& Ellingson 1997).
While there are legitimate concerns that large clusters
are not fully virialized, Carlberg et al. (1997a) have presented
spectroscopic evidence which strongly suggests that the clusters in the
CNOC survey are in equilibrium and, therefore, that the masses obtained
within the virial radius are reliable. 

Nevertheless,
because of its potential power to map the dark mass distribution
within a cluster independent of the cluster's dynamical state, recently
a great deal of effort has been devoted
to measurements of the gravitational potentials of clusters
via observations of weak lensing (e.g.\ Tyson, Wenk \& Valdes 1990;
Bonnet et al.\ 1994; Dahle, Maddox \& Lilje 1994; Fahlman et al.\ 1994;
Mellier et al.\ 1994; Smail et al.\ 1994, 1995, 1997; 
Tyson \& Fischer 1995; Smail \& Dickinson 1995; Kneib et al\ 1996;
Seitz et al.\ 1996; Squires et al.\ 1996ab; Bower \& Smail 1997;
Fischer et al.\ 1997; Fischer \& Tyson 1997; Luppino \& Kaiser 1997).
Relatively few clusters have been studied in detail, but a consistent
picture of the dark mass distribution appears to be emerging from
the weak lensing investigations.  In particular, the center of mass
corresponds well with the center of the optical light distribution and
the smoothed light distribution traces the dark mass well.
The lensing-derived mass to light ratios vary from cluster to cluster
but bracket a broad range of $200h (M/L)_\odot$ to $800h (M/L)_\odot$,
with most of the clusters falling in the middle of the range.

The consistency of cluster masses obtained from independent methods such
as lensing, virial analyses, or X-ray data (assuming pressure supported
hydrostatic equilibrium) is very much in debate at this time.  Cluster mass
estimates obtained from observations of strong lensing can often exceed
the X-ray mass by a factor of 2 to 3 (e.g. Miralda-Escud\'e \& Babul 1995).
This particular discrepancy is likely due to the failure of the assumption
of hydrostatic equilibrium and Waxman \& Miralda-Escud\'e (1995) have thus
proposed the existence of
multiphase cooling flows in the centers of rich clusters.  Based on analyses
of weak lensing over considerably larger 
cluster radii, however, some studies conclude that there is
quite good agreement between the lensing and X-ray masses (e.g.\ Squires
et al.\ 1996ab; Smail et al.\ 1997) while others claim a significant
disagreement in which the lensing mass systematically
exceeds the X-ray mass (e.g.\
Fischer et al.\ 1997; Fischer \& Tyson 1997).  Additionally, the weak 
lensing mass estimate is often found to exceed the mass obtained
from the virial estimator by
a factor of order 2 (e.g.\ Fahlman et al.\ 1994; Carlberg et al.\ 1994;
Smail et al.\ 1997) but in some cases good agreement between these two 
independently derived masses is found (e.g.\ Fischer et al.\ 1997).

One of the troubles associated with a fair assessment
of independent cluster mass estimates
is that different techniques tend, out
of necessity, to be applied at different cluster radii (i.e. it
is not always possible to investigate the gravitational potential
over the entire cluster via one particular estimator 
simply due to lack of data on appropriate scales).  
Systematic biases inherent in any given mass estimator may well
be scale-dependent and amongst different estimators the form of the
scale dependence is likely to vary.  Therefore, a considerable effort will
be required in order to reconcile all of the outstanding discrepancies
amongst independent cluster mass estimates.

Few observational investigations have been able to place direct
constraints on the radial mass profiles of clusters to date.  
Bonnet et al.\ (1994)
detected a coherent weak lensing shear due to the cluster Cl0024+1654
out to a radius of $r \sim 1.5h^{-1}$~Mpc and from their observations they
showed that the underlying cluster mass profile was consistent both with
an isothermal profile and a steeper de Vaucouleurs profile.  Similarly,
Fischer \& Tyson (1997) found that the weak lensing shear field of
RXJ 1347.5-1145 yielded a density profile that was consistent with
isothermal.  Tyson \& Fisher (1995), however, found that the density profile
implied by weak lensing observations of A1689 was steeper than isothermal
on large scales ($200 h^{-1} {\rm kpc} \ls r \ls 1h^{-1} {\rm Mpc}$).
Similarly, Squires et al.\ (1996b) found that the density profile of A2390
implied by the weak lensing shear field was consistent with isothermal
on small scales ($r \ls 250h^{-1} {\rm kpc}$), but on larger scales was
better described by a profile steeper than isothermal.  In contrast to
the weak lensing results, however, Carlberg et al.\ (1997b) found that
the velocity dispersion profiles of the CNOC clusters gave rise to a mean
cluster mass profile that was fit very well by a Navarro, Frenk \& White
profile, i.e.\ shallower than isothermal at small radii and isothermal at
large radii (e.g.\ Navarro, Frenk \& White 1995, 1996, 1997).

In this investigation we use
high-resolution N-body simulations of rich clusters 
to investigate systematic trends in both the total cluster masses and 
the radial
mass profiles as derived from
three simple estimators: (1) the weak lensing shear
field under the assumption of an isothermal
cluster potential, (2) the dynamical mass obtained from the measured velocity
dispersion under the assumption of an isothermal
cluster potential, and (3) the classical virial estimator.  The simulated
clusters are very massive and thus do not constitute an average, unbiased
sample of objects.  They do, however, correspond to the largest clusters
likely to form in a standard CDM universe and are objects which would
certainly be detectable as weak gravitational lenses.  The N-body
simulations of the clusters are discussed in \S2 and the weak lensing and
dynamical properties of the clusters are discussed in \S3 together with the 
mass profiles obtained from the three estimators. A discussion
of the results is presented in \S4.

\section{The Numerical Clusters}

The Hierarchical Particle-Mesh (HPM) N-body code written by J. V. Villumsen
(Villumsen 1989) was used to simulate the formation of three rich
clusters.  The HPM code allows small-volume particle-mesh
simulations to be nested self-consistently within large-volume
particle-mesh simulations and by successively nesting many simulations within
each other it is possible to obtain extremely high resolution in
both mass and length within a small, localized region of a large
computational volume (a ``power zoom'' effect).  
The code is, therefore, especially useful for the simulation
of the formation of objects such as individual clusters.  In particular,
using the HPM code to simulate the formation of clusters
obviates the need for computations which utilize
``constrained initial conditions'' and those which simulate at
high resolution the evolution of
density peaks that have been excised from the initial
conditions of a large computational volume.  While the largest-volume,
lowest-resolution
grid in an HPM simulation uses periodic boundary conditions,
the smaller-volume, higher-resolution grids use isolated boundary
conditions, allowing mass to flow in and out of the higher-resolution
grids over the course of the simulation.  
Therefore,  unlike constrained initial conditions or
peak-excision simulations, clusters simulated with
HPM are guaranteed to accrete all of the mass that they should
accrete if one simply ran a single large-volume simulation at
a level of resolution comparable to that of the highest-resolution
HPM grid.

The HPM code uses a standard cloud-in-cell (CIC) interpolation scheme,
which results in an approximately Gaussian smoothing of the power
spectrum with a smoothing length of $r_s = 0.8$ grid cell (see, e.g.,
\S6.6 of Blandford et al.\ 1991).  The gravitational force is, therefore,
softer than Newtonian on small scales but becomes Newtonian
for length scales greater than or of order 2 grid cells (Villumsen 1989).
Due to the force softening we therefore restrict our analyses to
length scales greater than 2 grid cells.

The clusters used for the present analysis are discussed in
detail in Brainerd, Goldberg \& Villumsen (1998) and here we present
only a summary of the simulations involved.  A standard Cold
Dark Matter model ($\Omega_0 = 1$, $\Lambda_0 = 0$, and $H_0 = 50$
km/s/Mpc) was adopted and the present epoch (i.e. redshift, $z$, of
0) was taken to correspond to $\sigma_8 = 1$ where
\begin{equation}
\sigma_8 \equiv 
\left< \left[ \frac{\delta\rho}{\rho}(8h^{-1} {\rm Mpc})
\right]^2 \right>^\frac{1}{2} .
\end{equation}
This is a model which is somewhat under-normalized compared to the
COBE observations (e.g.\ Bunn \& White 1997) and over-normalized
compared to the abundance of rich clusters (e.g.\ Bahcall \& Cen 1993;
White, Efstathiou \& Frenk 1993; Eke, Cole \& Frenk 1996;
Viana \& Liddle 1996). 
The simulations began at $\sigma_8 = 0.033$ (corresponding to a
redshift of 29) and were evolved forward in time to $\sigma_8 = 1.0$.

The formation of the three most massive (i.e.\ ``richest'') clusters 
contained within a single cubical volume of comoving side length $L = 400$~Mpc 
was followed at high resolution.
The large, primary computational
volume common to all three clusters was a standard particle-mesh
simulation consisting of $256^3$ grid cells and $128^3$ particles. 
Each cluster in turn was simulated at increased resolution by nesting
two smaller, higher-resolution grids successively within the
large, primary simulation volume.  
These two small, higher-resolution grids were 
centered on the center of mass of the particular cluster being simulated,
used $256^3$ grid cells each and had comoving
side lengths of $L = 66.6$~Mpc and $L = 16.7$~Mpc, 
respectively.
Since isolated boundary conditions are used in the small-volume simulations,
the number of particles physically inside the small grids 
varied over the course of the simulations.  

The particle mass in
the large computational volume common to all three clusters was
$m_p = 2.1\times 10^{12} M_\odot$, 
while for the smaller grids unique to the individual
clusters,  the particle masses were $m_p = 7.8\times 10^{10} M_\odot$
and $m_p = 9.8\times 10^9 M_\odot$ 
in the grids with $L = 66.6$~Mpc and
$L = 16.7$~Mpc, respectively.  Dynamic ranges
of $\sim 4.5\times 10^8$ in mass and $\sim 6000$ in length were thus
achieved by nesting the smaller simulations within the primary
computational volume.

Throughout the present analysis we shall define the clusters to consist
of all particles within a radius $r_{200}$ of the centers of mass,
where $r_{200}$ is the 
radius inside which the mean interior mass overdensity
is 200:
\begin{equation}
{\delta\rho \over \rho} \left( r_{200} \right) = 200
\end{equation}
(see, e.g., Navarro, Frenk \& White 1997, 1996, 1995).
The cluster mass estimates were computed using only particles 
from the three highest-resolution grids 
(i.e. the grids with $L = 16.7$~Mpc and $m_p = 9.8 \times 10^{9} M_\odot$)
at an epoch corresponding to a redshift of 0.5 ($\sigma_8 = 0.67$
for our normalization).   All of the particles within $r_{200}$ of
the cluster centers were excised from the highest-resolution grid
and were then used to compute the mass profiles.  
The total number of particles
per cluster 
located within $r_{200}$ at $z=0.5$ are: 192346
(``cluster 1''; $r_{200} = 2.1$~Mpc, proper radius), 288641 
(``cluster 2''; $r_{200} = 2.4$~Mpc, proper radius), and 310310 
(``cluster 3''; $r_{200} = 2.5$~Mpc, proper radius).
Within $r_{200}$ the clusters contain a significant amount of substructure
and have median projected ellipticities of 0.3. Cluster 1 is nearly prolate
while clusters 2 and 3 are nearly oblate.  Although the density profiles of
the clusters are fit well by Navarro, Frenk, \& White (1997, 1996, 1995)
profiles, the values of the
best-fit concentration parameters obtained for
the clusters are a factor of order 2 lower than the values predicted
by the Navarro, Frenk \& White formalism for objects in the identical
mass range.
For a full discussion of the above cluster properties see
Brainerd, Goldberg \& Villumsen (1998).

\section{Results}

Two-dimensional projections of the clusters are shown in the top
panels of Figs.\ 1, 2, and 3. 
The color scale shows the logarithm
of the surface mass density (in units of $M_\odot/{\rm kpc}^2$)
and distances are
given in proper coordinates for $z=0.5$.  
The actual mass profiles
of the clusters are shown in Fig.\ 4, where the top panel shows the
mean 2-dimensional projected mass profile obtained from 10 random 
projections of
each cluster and the bottom panel shows the 3-dimensional mass
profile of each cluster.  Throughout we shall adhere to notation in
which $R$ refers to a proper radius projected on the sky and
$r$ refers to a 3-dimensional proper radius.  In this notation, then,
$M(R)$ is the projected mass interior to a radius $R$ on the sky and $M(r)$ is
the mass interior to a sphere of radius $r$.
As expected from the work by Navarro, Frenk \& White (1997, 1996, 1995)
on the relatively generic shapes of the density profiles of objects formed
by dissipationless collapse, the mass profiles of the numerical clusters are
not fit well by single power laws (see also Dubinski \& Carlberg 1991,
Cole \& Lacey 1996, and Tormen, Bouchet \& White 1997).  
Rather, a gently changing slope
is observed, with the density profiles becoming roughly isothermal
on large scales ($\gs 1$~Mpc in the case of our clusters).

\vspace*{-2.0truecm}
\hbox{~}
\centerline{\psfig{file=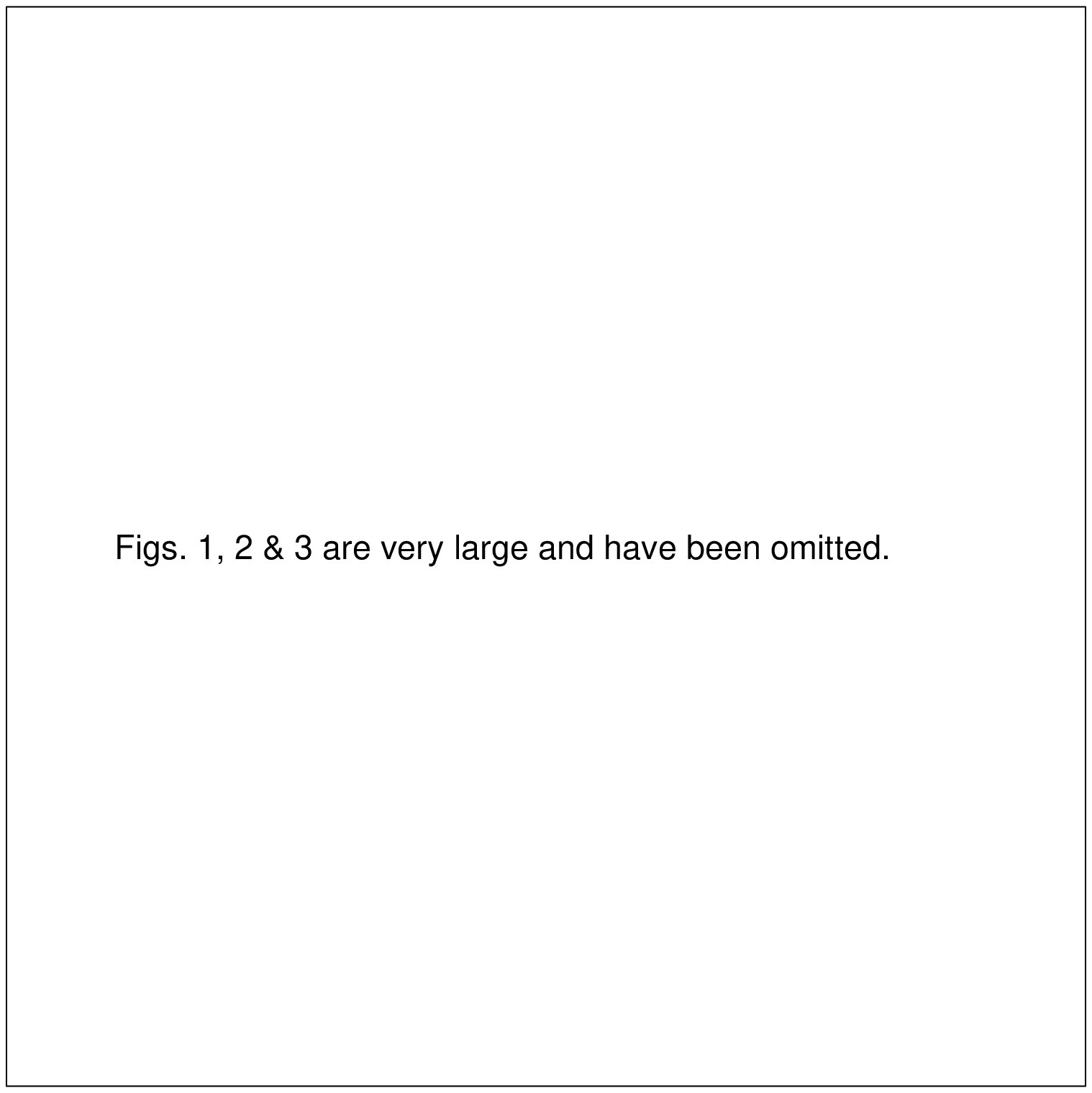,angle=0,width=3.0in}}
\vspace*{-2.truecm}
\noindent{\scriptsize
\addtolength{\baselineskip}{-3pt}
Fig.~1:
Top panel: the logarithm of the surface mass density
of cluster 1 as observed from a randomly-chosen line of sight.  The
units of the surface mass density are $M_\odot/{\rm kpc}^2$.  The cluster
consists of all particles in the highest-resolution subgrid that
are located within a radius $r_{200}$ of the center of
mass.
Bottom panel: the
gravitational lensing shear, $\gamma$, obtained for the projected mass density
shown in the top panel.  The cluster was placed at a redshift of
0.5 and the shear that would be induced in a plane of sources at $z=1.0$ was
computed by tracing a regular grid of $4\times 10^6$
light rays through the cluster.  The color scale
indicates the local value of $\log_{10} \gamma$
while the orientation of the sticks indicates the orientation
of the shear, $\varphi$.  For clarity, the mean orientation of the local
shear is
shown on a coarse $10 \times 10$ grid.  The angular scale of the figure
is of order $11' \times 11'$.
Figs. 2 and 3 are the same as Fig. 1, but for clusters 2 and 3, respectively.
\label{fig1}
\addtolength{\baselineskip}{3pt}
}

In the following subsections we will compare the true mass profiles
of the clusters to the mass profiles obtained from the three 
estimators.  All of the mass estimators assume the clusters to be
spherically symmetric
and, additionally, both the weak lensing and ``isothermal'' dynamical
mass estimates assume that the cluster potential is approximately isothermal.

\vspace*{-1.0truecm}
\hbox{~}
\centerline{\psfig{file=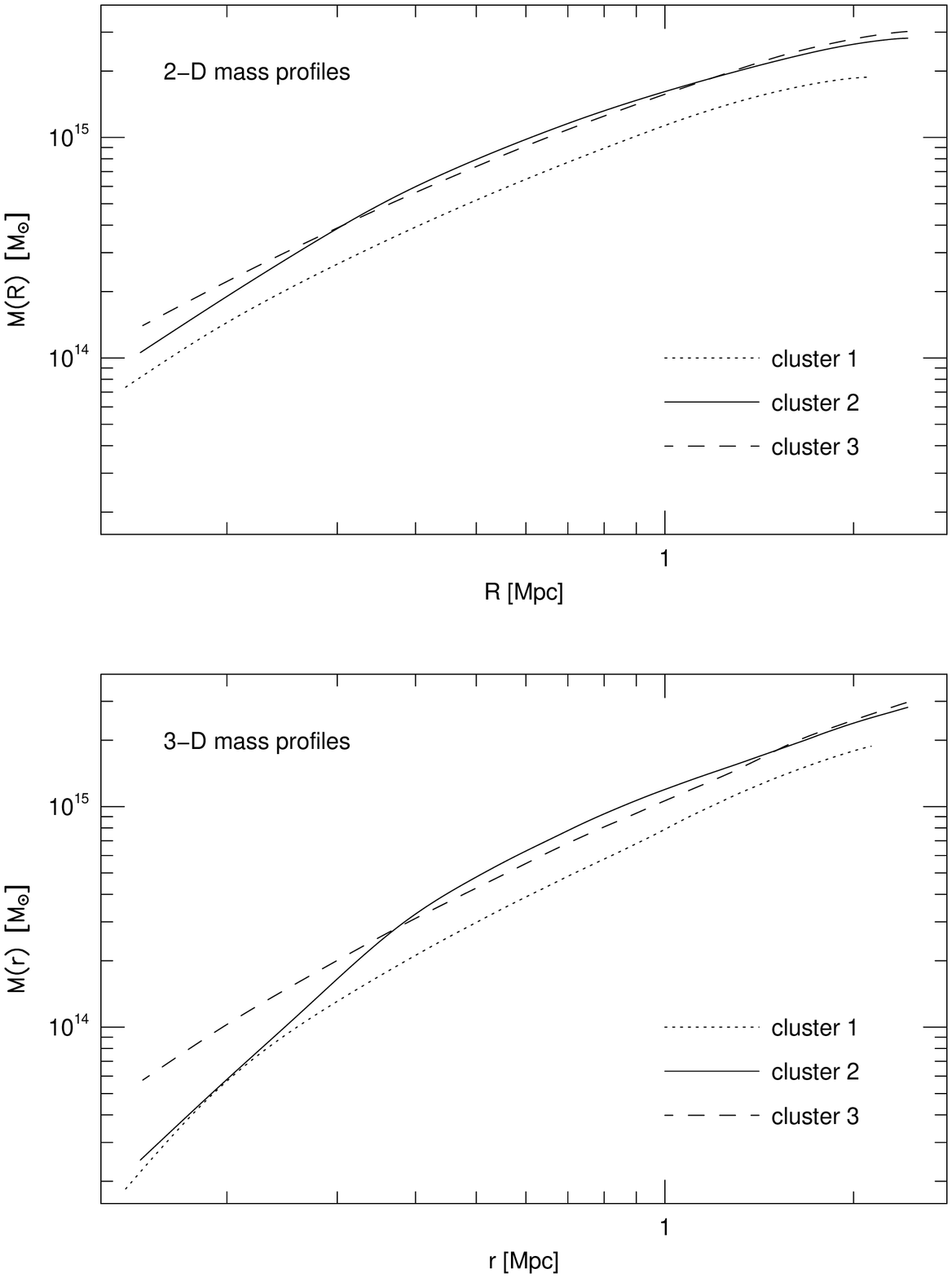,angle=0,width=4.5in}}
\vspace*{-1.0truecm}
\noindent{\scriptsize
\addtolength{\baselineskip}{-3pt}
Fig.~4:
The mass profiles of the clusters as computed directly from
the distribution of particles in the highest-resolution grids.  Top
panel: the mean projected mass profile, computed from 10 random
projections of each cluster.  Bottom panel: the full 3-dimensional mass
profile. The clusters are roughly isothermal on scales $\gs 1$~Mpc.
\label{fig4}
\addtolength{\baselineskip}{3pt}
}

\subsection{Weak Lensing Shear}

Observations of gravitational lensing provide potentially powerful
constraints on both the total mass and the mass distribution within clusters
of galaxies.  The gravitational potential of the cluster systematically
deforms the shapes of distant source galaxies that are seen through the
lensing cluster. The result is a net ellipticity induced in
the images of lensed galaxies and a net tangential alignment of 
the lensed images relative to the center of the cluster potential.

Provided the distance traveled by the light ray is very much
greater than the scale size of the lens itself, it is valid to
adopt the ``thin lens approximation'' in order to describe
a gravitational lens.  Consider a lens with an arbitrary
3-dimensional potential, $\Phi$.  In the thin lens approximation
a conveniently scaled 2-dimensional potential, $\psi$, is adopted for
the lens (i.e.\ $\psi$ is a scaled representation of
the 3-dimensional potential of the lens integrated
along the optic axis):
\begin{equation}
\psi(\vec{\theta}) = \frac{D_{ds}}{D_d D_s} \; \frac{2}{c^2}
\int \Phi(D_d \vec{\theta},z) \;\; dz .  
\end{equation}
Here $\vec{\theta}$ is the location of the lensed image on the
sky relative to the optic axis, $D_{ds}$ is the angular diameter 
distance between the lens (the ``deflector'') and the source, $D_d$
is the angular diameter distance between the observer and the lens, and
$D_s$ is the angular diameter distance between the observer and the source.

Having adopted this 2-dimensional lens potential, then, it is
straightforward to relate the potential of the
lens (through second derivatives of $\psi$) directly to
the two fundamental quantities which
characterize the lens: the convergence ($\kappa$) and
the shear ($\vec{\gamma}$).  The convergence, which
describes the isotropic focusing of light rays, is given by:
\begin{equation}
\kappa(\vec{\theta}) = \frac{1}{2} \left( \frac{\partial^2 \psi}
{\partial \theta_1^2} +  \frac{\partial^2 \psi}{\partial \theta_2^2}
\right) .  
\end{equation}
The shear describes the tidal gravitational forces acting across
a bundle of light rays and, therefore, the shear
has both a magnitude, $\gamma = \sqrt{\gamma_1^2 + \gamma_2^2}$,
and an orientation, $\varphi$.  In
terms of $\psi$, the components of the
shear are given by:
\begin{equation}
\gamma_1 (\vec{\theta}) = \frac{1}{2} \left( \frac{\partial^2 \psi}
{\partial \theta_1^2} - \frac{\partial^2 \psi}{\partial \theta_2^2}
\right) \equiv \gamma(\vec{\theta}) \cos \left[ 2\varphi(\vec{\theta}) \right]
\end{equation}
\begin{equation}
\gamma_2 (\vec{\theta}) 
= \frac{\partial^2 \psi}
{\partial \theta_1 \partial \theta_2}
= \frac{\partial^2 \psi}{\partial \theta_2
\partial \theta_1} \equiv \gamma(\vec{\theta}) \sin \left[
2 \varphi(\vec{\theta}) \right] 
\end{equation}
(e.g.\ Schneider, Ehlers \& Falco 1992).

A great deal of work has been done in recent years to develop methods
by which a map of the surface mass density of a cluster can be 
reconstructed from observations of the distortions induced in the images
of background galaxies in the limit of weak gravitational lensing, for which
$\kappa << 1$ and $|\gamma| << 1$ (e.g.\ Kaiser \& Squires 1993;
Bartelmann 1995; Kaiser 1995; Kaiser et al.\ 1995; Schneider 1995; 
Schneider \& Seitz 1995; Seitz \& Schneider 1995; Bartelmann et al.\ 1996;
Seitz \& Schneider 1996; Squires \& Kaiser 1996; Seitz et al.\ 1998).  
It is not the intent
of this paper to explore these detailed methods of cluster mass
reconstruction.  Rather, we will
focus on a very simple weak lensing analysis technique
that is sometimes used 
to gauge the total mass of a cluster contained within
a given radius without fully reconstructing the underlying density profile.

The method invokes an assumption that the cluster potential may
be represented adequately by an isothermal sphere.  The actual density
potentials of the numerical clusters
are better represented by Navarro, Frenk \&
White profiles (see Brainerd, Goldberg \& Villumsen 1998) than by singular
isothermal spheres and, given the apparent generality of the NFW profile,
it is more likely that an NFW profile will better represent an actual
galaxy cluster than will an isothermal sphere.   Here we choose to adopt
the isothermal sphere approximation for the analysis because this is
the simplifying assumption that is most commonly invoked in the literature
when cluster masses are estimated from observations of weak lensing without
a full reconstruction of the density profile (see the references
listed below).  Here our goal is simply to quantify systematic effects due
to the assumption of an underlying isothermal potential when the true
potential is better approximated by that of an NFW-type object.

An isothermal
sphere is uniquely specified by a single quantity, the velocity
dispersion ($\sigma_v$), and the mass of an isothermal sphere contained within
a 3-dimensional radius $r$ is given by
\begin{equation}
M(r) = \frac{2 \sigma_v^2 r}{G}
\end{equation}
where $G$ is Newton's constant.  The total mass of an isothermal sphere
within a radius $R$ projected on the sky is given by
\begin{equation}
M(R) = \frac{\pi \sigma_v^2 R}{G} 
\end{equation}
(e.g.\ Binney \& Tremaine 1987). 

Since it is spherically symmetric, the isothermal
sphere gives rise to a gravitational lensing shear field which is necessarily
circularly symmetric and, in particular, the
shear as a function of angular radius, $\theta$, is given by
\begin{equation}
\gamma(\theta) = \frac{2\pi}{\theta} \left( \frac{\sigma_v}{c} \right)^2 \left[
\frac{D_{ds}}{D_s} \right] ,
\end{equation}
where $c$ is the velocity of light and $\sigma_v$ is the velocity
dispersion of the lens (e.g.\ Schneider, Ehlers \& Falco 1992).  
If we consider an annulus of
inner radius $\theta_{\rm min}$ and outer radius $\theta_{\rm max}$
centered on the center of mass of the isothermal sphere,
the mean shear inside the annulus is given by
\begin{equation}
\overline{\gamma} = 4\pi \left( \frac{\sigma_v}{c} \right)^2 \left[
\frac{D_{ds}}{D_s} \right]
\left( \theta_{\rm max} + \theta_{\rm min} \right)^{-1} .
\end{equation}
That is, provided the cluster potential is sufficiently well-represented
by an isothermal sphere it is possible to deduce its characteristic
velocity dispersion directly from either a measurement of the shear at a given
radius, $\gamma (\theta)$, or the mean value of the shear, $\overline{\gamma}$,
computed within some large
large annulus.  A measurement of
$\sigma_v$ by such a technique then leads to an estimate 
of the mass of the cluster within a given radius (e.g. 
Tyson, Wenk \& Valdes 1990; Bonnet et al.\ 1994; Smail et al.\ 1994,
1997; Smail \& Dickinson 1995;  Bower \& Smail 1997; Fischer \& Tyson 1997).

In practice, an observed weak lensing shear only places a limit
on the mass of cluster to within an additive constant (the
so-called uniform density mass sheet degeneracy).  The simple
singular isothermal sphere mass estimator that we use
here formally assumes that there is no such
mass sheet present and that the observed weak lensing shear can be directly
translated into a mass measurement via equations (7), (8), (9),
and (10).
Below we will, therefore, interpret the cluster shear fields in a
manner consistent with the simple form of the mass estimator and
we will not explicitly address the mass sheet degeneracy problem or
its implications for an observed weak lensing shear.

In this section we compute the shear fields of the numerical clusters and
in \S3.3 we will use these shear fields
to investigate the systematic effects that the above weak lensing
mass estimate has on the masses inferred for the numerical clusters.
The shear fields of the clusters are determined directly by tracing regular 
Cartesian grids
of $2001 \times 2001$ light rays through the clusters.  In
the analysis below
we adopt the thin lens approximation and for a particular plane
projection of a cluster we simply calculate the net
deflection of each light ray due to all of the point masses contained
within $r_{200}$ of the cluster center of mass.  
Note, however, that we ran a few test cases in which all particles inside
a radius of 4~Mpc of the cluster centers were included in the ray trace 
analysis. The inclusion of the mass exterior to a radius of
$r_{200}$ gave rise to a shear field interior to $r_{200}$ that was
indistinguishable from the shear field obtained using only the
particles interior to $r_{200}$.  That is, owing to the fact that the
clusters are roughly axisymmetric and no large mass concentrations 
exist just outside the clusters, the shear interior to a projected
radius $R$ is determined by the surface mass density interior
to $R$.

The clusters are located at a redshift of $z = 0.5$
and we consider a plane of sources at $z = 1.0$.  (Although
the redshift of the sources will affect the magnitude of the shear,
it will not affect the velocity dispersion inferred in the isothermal
sphere approximation and, therefore, the choice of source plane is
essentially arbitrary for our analysis.)  
The side lengths of the grids of light rays
were taken to be $L = 2r_{200}$ so that throughout we compute only
the shear interior to the virial radii of the clusters.  At the redshift
of the clusters, then, the side lengths of the grids correspond to 
and angular scale of order $11' \times 11'$.

If we let the location of a light ray on the grid
be given by $\vec{\beta}$ prior to
lensing (i.e. $\vec{\beta}$ is the location of the light ray in the source
plane) and we let $\vec{\theta}$ be the location of the light ray after
having been lensed by all of the point masses (i.e. $\vec{\theta}$ is the
location of the light ray in the image plane), the
components of the shear are then given by:
\begin{equation}
\gamma_1(\vec{\theta}) = - \frac{1}{2} \left( \frac{\partial 
\beta_{\rm x}}
{\partial \theta_{\rm x}} - \frac{\partial \beta_{\rm y}}
{\partial \theta_{\rm y}} \right)
\end{equation}
\begin{equation}
\gamma_2(\vec{\theta})= - \frac{1}{2} \left( \frac{\partial \beta_{\rm x}}
{\partial \theta_{\rm y}} + \frac{\partial \beta_{\rm y}}
{\partial \theta_{\rm x}} \right)
\end{equation}
The $2001 \times 2001$ light rays define a grid of $2000 \times 2000$ cells
and the shear at the centers of each of these cells can be determined
from equations (11) and (12) above by finite differencing of the deflections
of the four light rays which define the corners of the cell. 

The code used to compute the net deflections of the grid of light
rays was tested by tracing the light rays through a number of singular
isothermal spheres that were
approximated by a set 250,000 point masses.  The point
masses were constrained to lie within a maximum projected radius of $R=2.7$~Mpc
and their  masses were scaled appropriately so as to reproduce the
correct values of $M(R=2.7 \; {\rm Mpc})$ for a  set of isothermal spheres
with values of $\sigma_v$ in
the range of 500 km/s to 1500 km/s.
As with the simulated clusters, the isothermal sphere lenses were
placed at $z=0.5$ and the source light rays emanated from $z=1.0$.
The net deflections of the light rays were evaluated and the radial
dependence of the convergence, $\kappa(R)$, and shear, $\gamma(R)$,
was computed and compared to the analytic expectations for infinite
singular isothermal spheres having values of $\sigma_v$ identical to
the isothermal spheres that were approximated by the point masses.
For the isothermal sphere we know $\kappa(R) = \gamma(R)$, and in all cases
good agreement was found between both $\kappa(R)$ and $\gamma(R)$ as computed
individually from the ray tracing and between the ray tracing
results and the analytic expectations (deviations $\ls 1$\% of
the analytic values).

Shown in the bottom panels of Figs.\ 1, 2, and 3 are the shear fields
corresponding to the 2-dimensional
projections of the clusters shown in the
top panels of these figures.  The color scale shows the logarithm
of the magnitude of the
shear and the small sticks indicate its orientation. For clarity of
the figure, we plot 
the mean orientation of the shear 
on a coarse $10 \times 10$ grid 
that was computed from an unweighted average of the local shear vectors
obtained from the differencing of the displacements of the 
light rays (i.e. the sticks show a rebinning of the original 
$2000 \times 2000$ grid of shear vectors onto a $10 \times 10$ grid).
The visual agreement of the magnitude and orientation of the shear with
the actual surface mass density of the clusters is as expected; the
shear is greatest in the densest regions of the clusters and is oriented
roughly tangentially with respect to the cluster centers.  
The shear fields are not, however, circularly symmetric and reflect 
both the overall ellipticity of the clusters and the substructure
within them.

\vspace*{-1.3truecm}
\hbox{~}
\centerline{\psfig{file=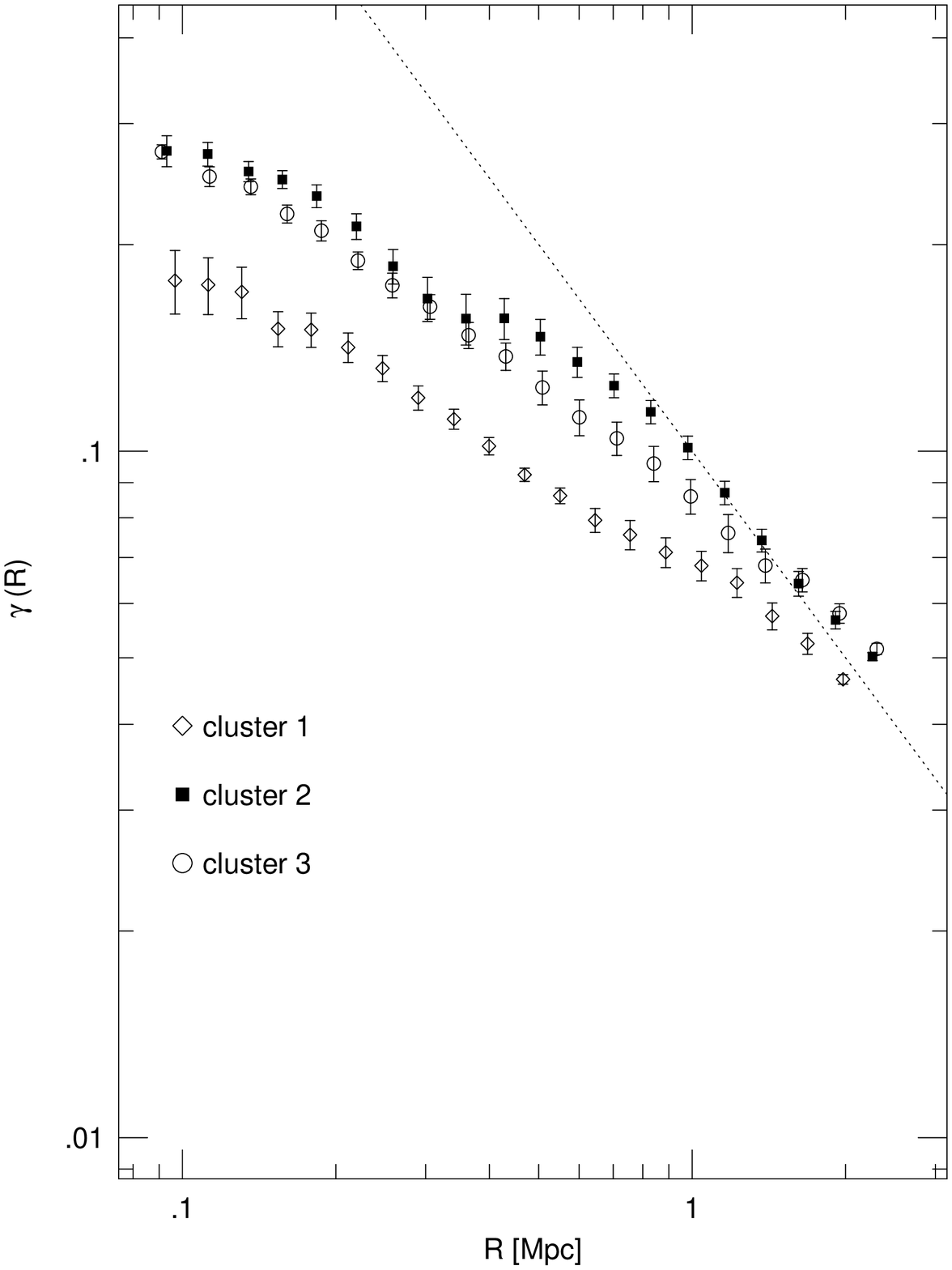,angle=0,width=4.0in}}
\vspace*{-1.0truecm}
\noindent{\scriptsize
\addtolength{\baselineskip}{-3pt}
Fig.~5:
The mean gravitational lensing shear for the clusters as a function
of projected radius.  Two-dimensional shear fields were determined for 10
random projections of each cluster, from which the average radial
value of the shear
was computed in independent bins of radius $R$ centered on the cluster
center of mass.  The error bars show
the formal standard deviation in the mean between the 10 projections.
For comparison the dotted line indicates the shape of the shear profile
expected for
an isothermal sphere lens, $\gamma(R) \propto R^{-1}$.
\label{fig5}
\addtolength{\baselineskip}{3pt}
}

Each cluster was viewed at 10 random orientations and a mean radial
shear profile was computed from the full $2000 \times 2000$ grid
of shear vectors.  The results are shown in Fig.\ 5,
where the error bars indicate the formal standard deviation in the
mean between the 10 random projections.  Also shown for comparison
is the radial shear profile expected for an isothermal sphere
(i.e.\ $\gamma (R) \propto R^{-1}$, cf.\ equation 9 above).  Below a
scale of $\sim 1$~Mpc the radial shear profiles of the clusters 
behave as $\gamma(R) \propto R^{-0.5}$, 
while on larger scales the variation
of $\gamma$ with $R$ is roughly isothermal, $\gamma(R) \propto R^{-1}$.
Given the mass profiles
shown in Fig.\ 4, this is precisely the behavior we would anticipate for
the shear profiles.  This behavior will, however, cause 
systematic errors in the cluster masses inferred from
the shear fields under the assumption of isothermal cluster potentials.

\subsection{Velocity Dispersions}

Under the assumption of an isothermal cluster potential, the masses 
of the clusters can be determined from measurements of 
their velocity dispersions
alone (e.g.\ equations 7 and 8 above).  The isothermal sphere is characterized
by a single, constant value for the velocity dispersion and in this
section we investigate the degree to which the measured cluster
velocity dispersions
vary with distance from the cluster centers of mass.
The force resolution of the simulations is too poor to resolve convincingly
the dark matter halos that would be associated with individual galaxies
within the cluster (one grid cell in the particle-mesh calculation
is of order 45 kpc in length) and,
so, it is not possible to calculate the line of sight velocity
dispersion of member galaxies directly.  However, in the absence of significant
velocity bias in both observed and high-resolution numerical
clusters (e.g.\ Lubin \& Bahcall 1993; Bromley et al.\ 1995;
Ghigna et al.\ 1998),
a random subset of the particles can be drawn from each cluster
to estimate the velocity dispersion that would be expected for the
member galaxies.

Each cluster was viewed from 1000 random orientations and the line of
sight velocity dispersion, $\sigma_v$, was computed as a function of
projected radius relative to the cluster center of mass.  
Two types of annuli were used for the computation:
independent annuli (i.e.\ $\sigma_v$ was computed in thin annuli with
differential radius, $R$) and
cumulative annuli (i.e.\ $\sigma_v$
was computed in wide annuli which shared a fixed inner radius, $R_{\rm min}$,
and differed only by the maximum radius of the annuli, $R_{\rm max}$).  
That is, the use of the independent annuli yields a measurement of
$\sigma_v$ at a particular projected distance from the cluster center
while the use of the cumulative annuli yields a measurement of 
$\sigma_v$ averaged over the entire cluster (out to some maximum radius).
Throughout, the minimum radius from the cluster centers of mass,
$R_{\rm min}$, was taken to be a distance equal to
the length of two grid cells in the particle-mesh simulation since
below that scale the gravitational force is softened by the
N-body computational technique. 

Shown in Figs.\ 6 and 7 (crosses)
are the mean values of $\sigma_v$ that were
calculated directly from the line of sight velocities of
particles within the clusters.  Independent annuli were
used in Fig.\ 6 and cumulative annuli were used in Fig.\ 7.  The error
bars in the figure show the formal 1-$\sigma$ dispersion amongst the
different projections of the clusters.  The velocity dispersion computed
using independent annuli decreases monotonically with radius in all
three clusters but the decrease is slow enough such that averaged over
large scales within the clusters (i.e. $\sigma_v$ computed in the
cumulative annuli) the velocity dispersion is roughly constant.

\vspace*{-0.7truecm}
\hbox{~}
\centerline{\psfig{file=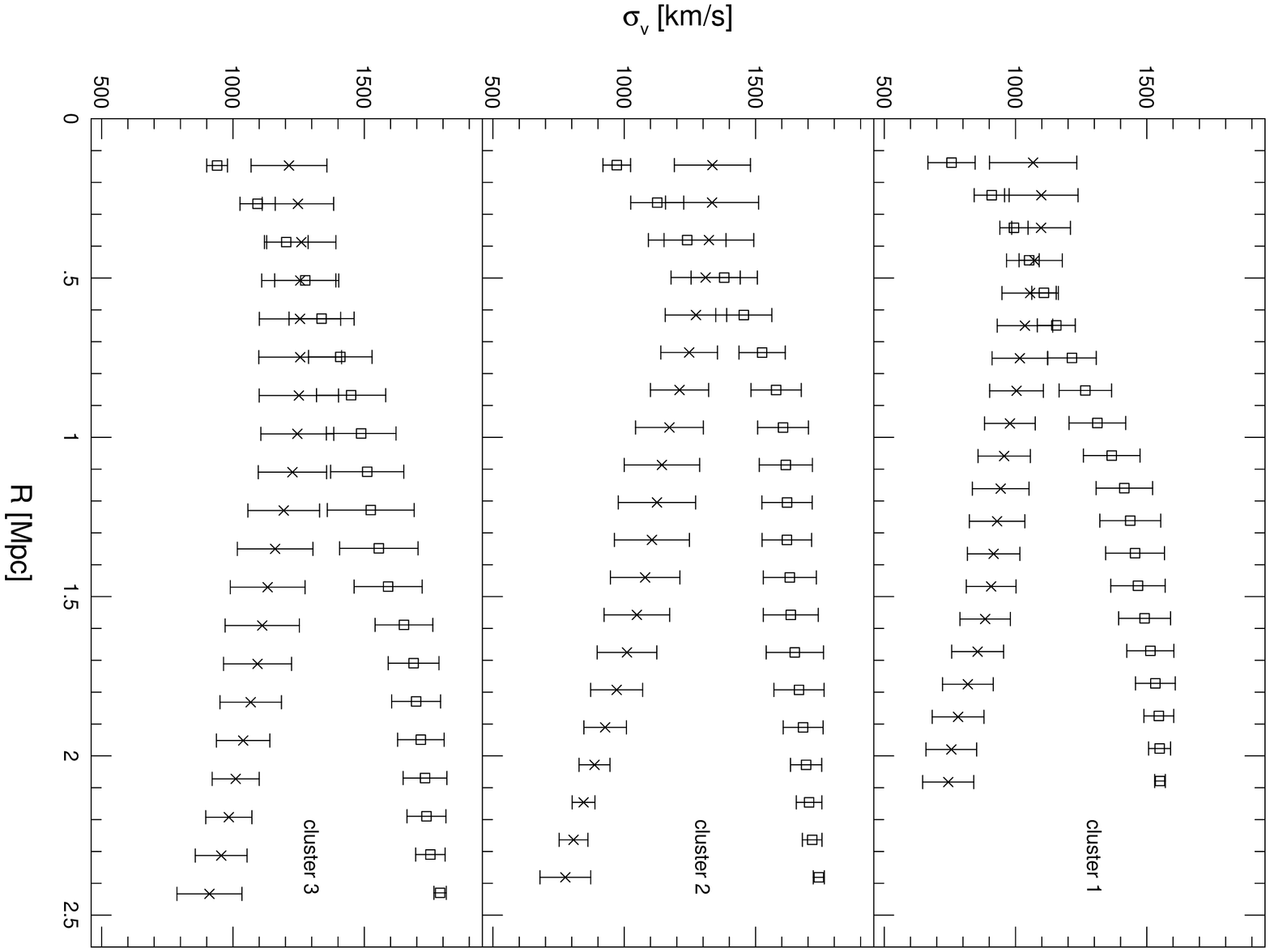,angle=270,width=4.5in}}
\vspace*{1.7truecm}
\noindent{\scriptsize
\addtolength{\baselineskip}{-3pt}
Fig.~6:
Line of sight cluster velocity dispersions, $\sigma_v$, as function
of projected radius.  Crosses show the mean value of $\sigma_v$ computed
directly from the velocities of
random subsets of the constituent particles and the error bars
show the formal 1-$\sigma$ deviation amongst 1000 random lines of sight.
Squares show the value of $\sigma_v$ inferred for the clusters on the
basis of the mean weak lensing shear, under the assumption that the
cluster potential is well-represented by an isothermal sphere;  error
bars show the formal 1-$\sigma$ deviation amongst the 10 random lines of
sight for which direct ray tracing was performed.
In this figure $\sigma_v$ has been computed using independent annuli
with proper radius $R$.
\label{fig6}
\addtolength{\baselineskip}{3pt}
}

Also shown in Figs.\ 6 and 7 (squares) are the mean values of $\sigma_v$
that are
{\it inferred} for the clusters on the basis of the weak lensing 
shear field, assuming that the cluster potentials can be
well-represented by isothermal spheres (e.g.\ Fig.\ 5).  
From the 10 different 
projections for which direct ray tracing was performed, the
mean shear was computed using both
independent and cumulative annuli identical to the annuli used to
compute the velocity dispersions of the particles themselves.  The values of
$\gamma(R)$ and $\overline{\gamma}(R_{\rm max})$ obtained from the ray
trace analysis were then used in conjunction with equations (9) and (10)
to infer the variation of the cluster velocity 
dispersion with radius.  Error bars in Figs.\ 6 and 7 show
the formal 1-$\sigma$ dispersion amongst the different cluster projections.
In contrast to the velocity dispersion measured directly for the particles,
the velocity dispersion inferred from the weak lensing analysis increases
with radius monotonically.

%\vspace*{-4.0truecm}
\hbox{~}
\centerline{\psfig{file=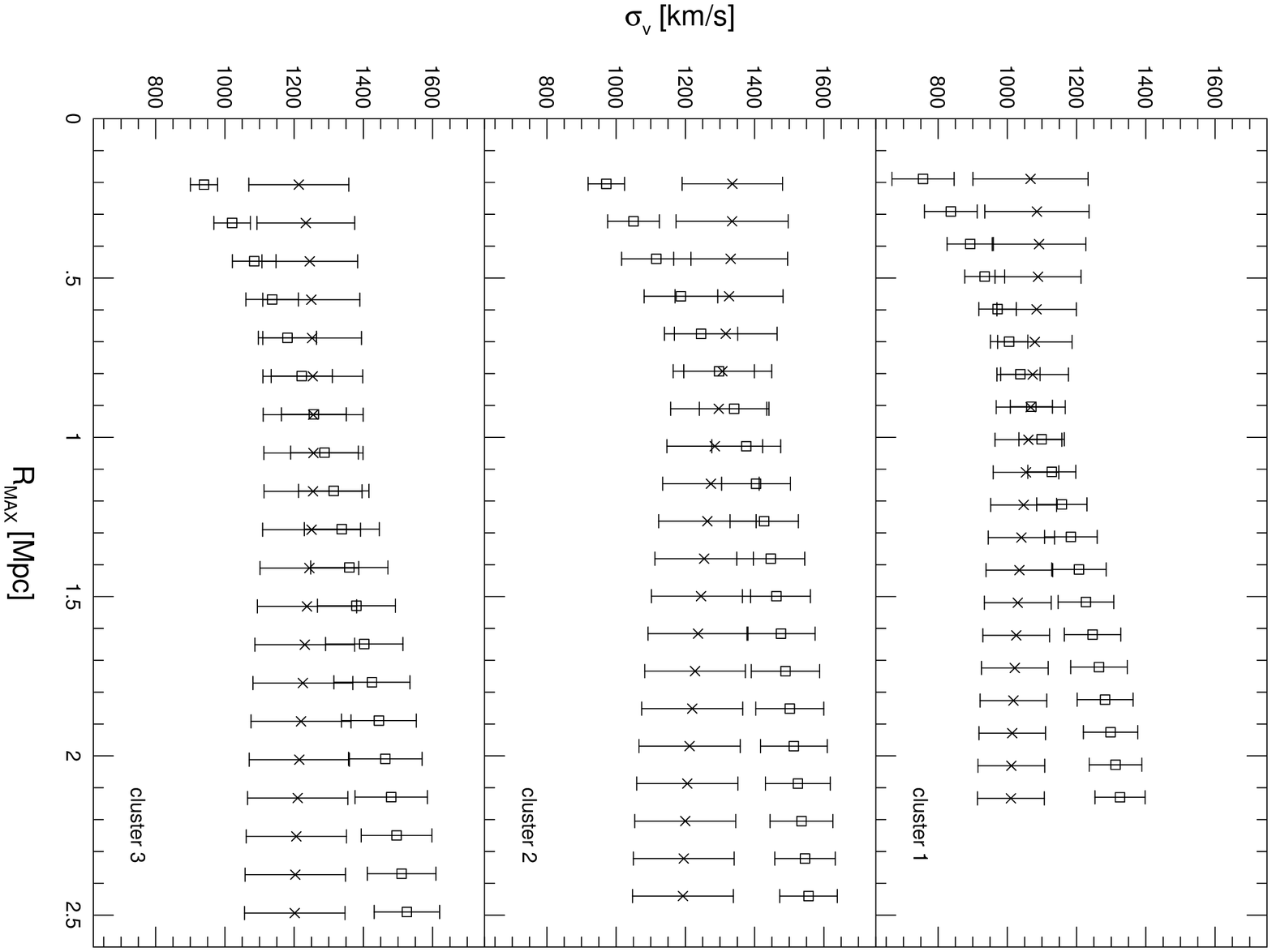,angle=270,width=4.5in}}
\vspace*{1.7truecm}
\noindent{\scriptsize
\addtolength{\baselineskip}{-3pt}
Fig.~7:
Same as Fig.\ 6 except that in this figure
$\sigma_v$ has been computed
using large cumulative radii of outer radius $R_{\rm max}$ (see text).
\label{fig7}
\addtolength{\baselineskip}{3pt}
}

\subsection{Cluster Mass Estimates}

Here we compute mass profiles for the clusters using the following simple
estimators: (1) the mean value of the weak lensing shear under the
assumption of an isothermal cluster potential, (2) the dynamical mass
obtained from the line of sight velocity dispersion of the particles under the
assumption of an isothermal cluster potential, and (3) the classical
virial estimator.  The cluster mass profiles obtained using the estimators
are compared directly to the true mass profiles (e.g.\ Fig.\ 4) and
throughout we will plot ratios of the estimated and 
true cluster mass profiles as a function of radius.

Shown in Figs.\ 8 and 9 are the mass profiles obtained from the mean
weak lensing shear under the assumption of an isothermal cluster
potential.  Results for the 2-dimensional projected mass profiles
are shown in Fig.\ 8 and the 3-dimensional mass profiles are shown
in Fig. 9.  The velocity dispersion, $\sigma_v(R)$, inferred from
the circularly-averaged weak lensing signal (e.g.\ the squares in
Fig.\ 6 and 7) was used 
in equations (7) and (8) above to compute $M(r)_{\rm lens}$ and 
$M(R)_{\rm lens}$.  In both
Figs.\ 8 and 9 the circles indicate that the value of $\sigma_v(R)$
was determined using the large, cumulative annuli (i.e.\ an average
velocity dispersion over the cluster out to a maximum radius of $R$).  
The solid squares
in these figures indicate that the value of $\sigma_v(R)$ was determined
using thin, independent annuli (i.e.\ a value of the velocity dispersion
computed at a particular distance, $R$, from the cluster center of
mass).  From the weak lensing analysis it is
not possible to measure the direct dependence of $\sigma_v$ on the
3-dimensional radius, $r$, and, so, to compute the 3-dimensional mass
profile we have taken the velocity dispersion to be
$\sigma_v(r) \equiv \sigma_v(R=r)$.

\vspace*{-1.0truecm}
\hbox{~}
\centerline{\psfig{file=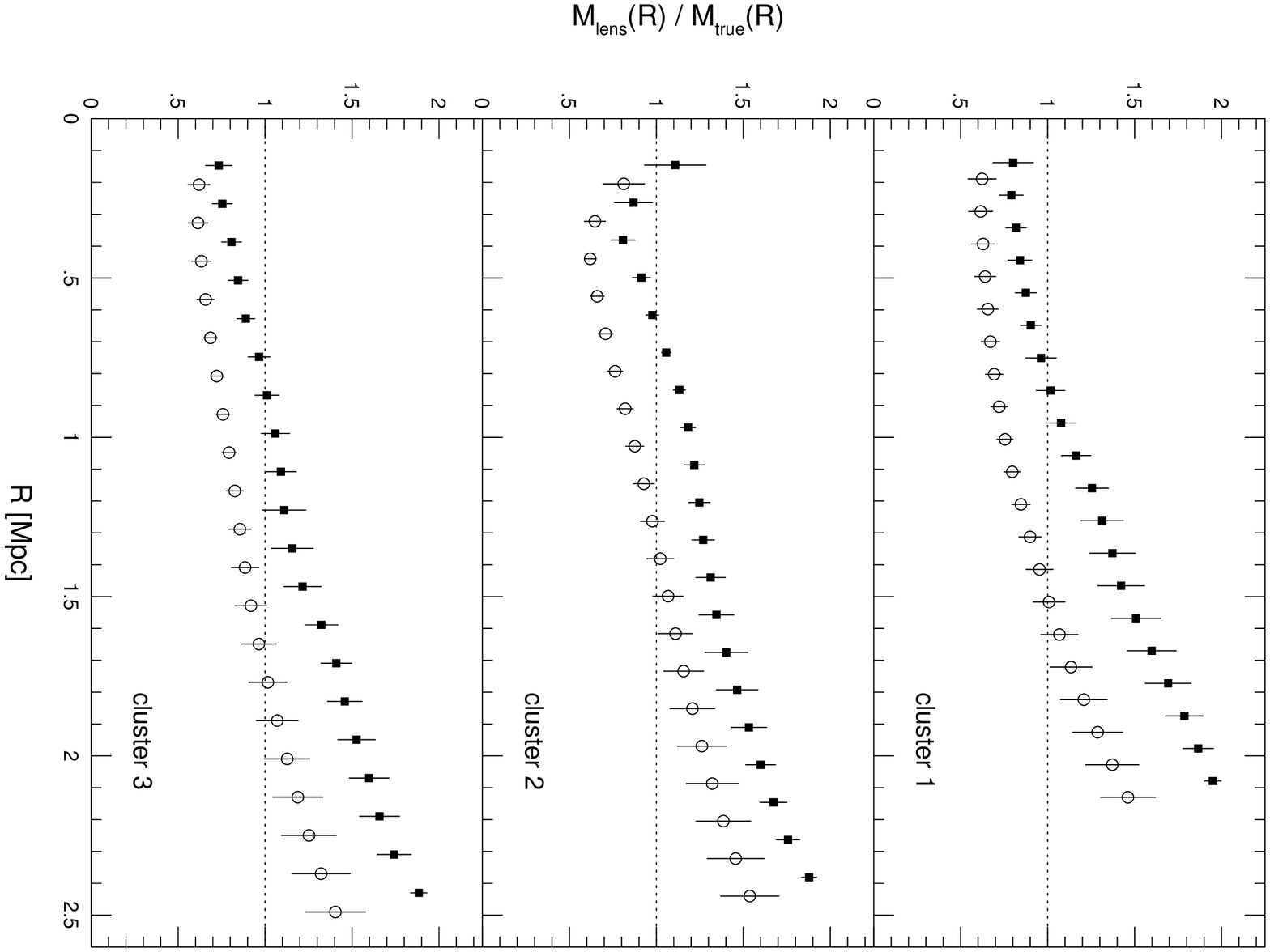,angle=270,width=4.5in}}
\vspace*{1.7truecm}
\noindent{\scriptsize
\addtolength{\baselineskip}{-3pt}
Fig.~8:
The 2-dimensional, projected cluster mass profile
obtained from the weak lensing analysis compared to the
true cluster mass profile.
Solid squares indicate that the value of $\sigma_v$ used in equation (8)
was determined from independent annuli of differential radius $R$;
open circles indicate that the
value of $\sigma_v$ was determined from large, cumulative annuli with
outer radii of $R_{\rm max} = R$.
The data points shown by the squares have been plotted such that $R$ is
the value of the projected radius at the midpoints of the independent
radial bins
and the data points shown by the
circles are plotted such that $R$ is the value of $R_{\rm max}$
(i.e.\ for the circles
$R$ corresponds to the outermost radius of the annulus used in
the calculation).
Error bars show the 1-$\sigma$ dispersion
in $M(R)$
amongst the 10 different projections for which ray tracing was performed.
\label{fig8}
\addtolength{\baselineskip}{3pt}
}

\bigskip
Fig.\ 8 shows that there is clearly a scale-dependent systematic deviation
of the 2-dimensional projected cluster mass profile determined from the
simple weak lensing analysis adopted here.
Overall the trend is for $M(R)_{\rm lens}$ to increase monotonically
with radius, underestimating the true projected mass at small radii and
overestimating the true projected mass at large radii.  The overestimate
of the projected mass at large radii is simply a reflection of the fact that
the isothermal sphere is, by definition, infinite in extent while the
actual clusters are confined to a finite radius of $r_{200}$.  (Note,
however, that we performed a few test cases in which
the proper radius of the numerical clusters was increased to a value  of
$r=4$~Mpc and this had a negligible effect upon the
measured shear and, hence, the inferred projected mass.) 

In contrast to
the results for $M(R)_{\rm lens}$, there is only a weak scale
dependence in the deviation of $M(r)_{\rm lens}$ from the true 3-dimensional
mass.  Over most scales there is quite good agreement between
the true cluster mass profiles and $M(r)_{\rm lens}$ as determined from 
values of $\sigma_v$ that were
computed using independent annuli.
When values of $\sigma_v$ determined from the
large cumulative annuli are used, $M(r)_{\rm lens}$ systematically
underestimates the true cluster mass on scales significantly less than
$r_{200}$.  At large radii, however, $M(r)_{\rm lens}$ is in very good
agreement with the true mass of the cluster for the case in which 
$\sigma_v$ was determined from the large, cumulative annuli (i.e. 
$\sigma_v$ is determined from the mean shear over the entire cluster).
This result may seem a bit surprising given the fact that the
clusters are better represented by NFW density profiles than they are by
isothermal
spheres. However, for NFW-type objects with masses comparable to those
of our numerical clusters, the mean shear interior to $R_{200}$  differs
relatively little ($\ls 10$\%) from that of an isothermal sphere that has
an identical mass contained within $r_{200}$ (Oaxaca Wright \& Brainerd 1999).
Hence, the isothermal sphere approximation should yield a reasonable
estimate of the cluster mass contained within $r_{200}$, provided the
mean shear is computed interior to a projected radius of $R_{200}$. 

\vspace*{-1.0truecm}
\hbox{~}
\centerline{\psfig{file=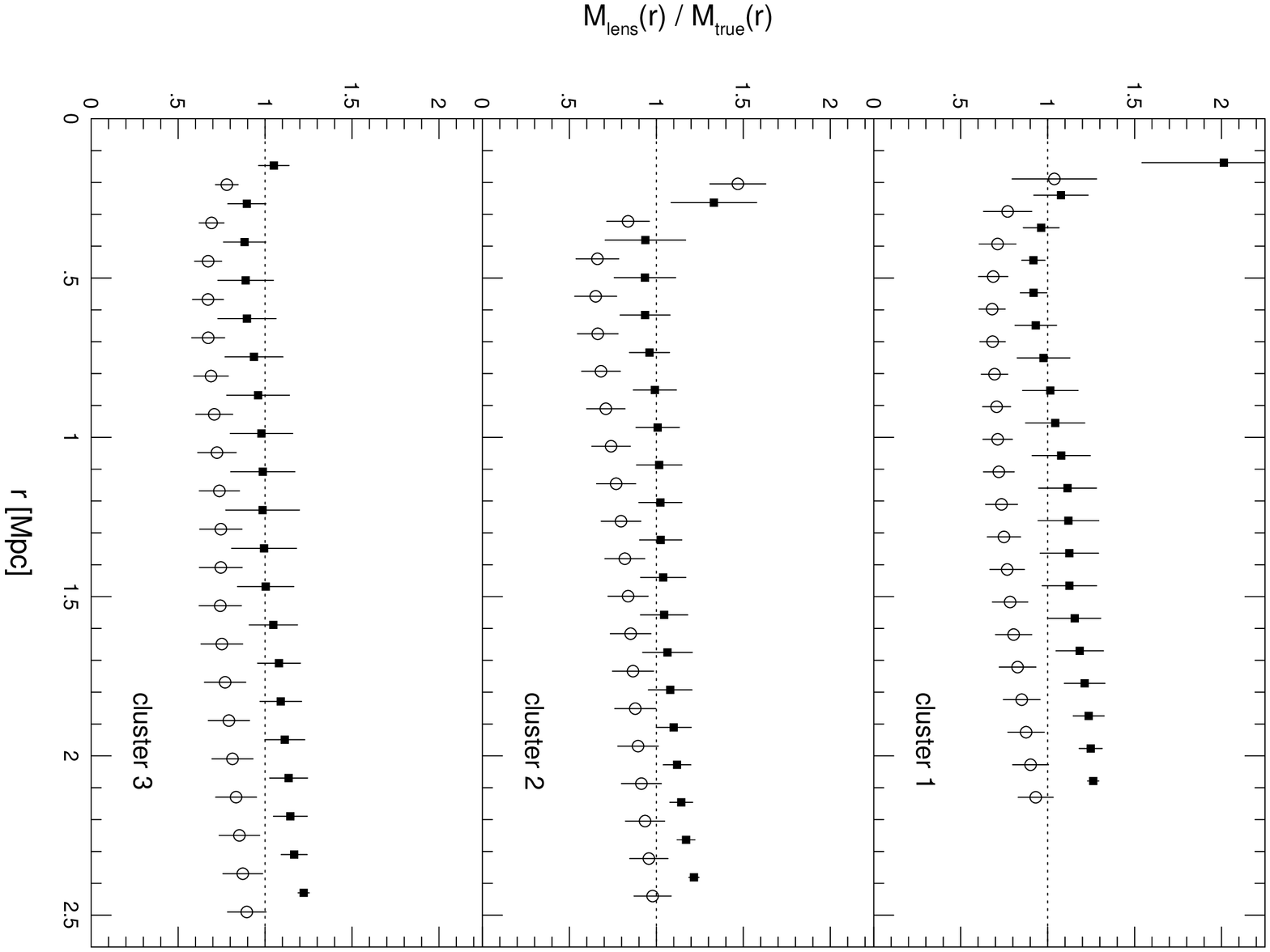,angle=270,width=4.5in}}
\vspace*{1.7truecm}
\noindent{\scriptsize
\addtolength{\baselineskip}{-3pt}
Fig.~9:
The 3-dimensional cluster mass profile obtained from
the weak lensing analysis compared to the true cluster mass profile.
Solid squares indicate that the value of $\sigma_v$ used in equation (7)
was determined from independent annuli of differential radius $R$;
open circles indicate that the value of $\sigma_v$ was determined from
large, cumulative annuli with $R_{\rm max} = r$.  The data points shown
by the squares have been plotted such that $r$ is the value of the
3-dimensional radius at the midpoints of the independent radial bins
and the data
points shown by the circles are plotted such that $r$ is the value
of $r = R_{\rm max}$.
Error bars show the
1-$\sigma$ dispersion in $M(r)$ amongst the 10 different projections for which
ray tracing was performed.
\label{fig9}
\addtolength{\baselineskip}{3pt}
}

Shown in Figs.\ 10 and 11 are the cluster mass profiles obtained from
equations (7) and (8) in which $\sigma_v$ is taken to be the mean particle 
velocity dispersion measured
directly from random subsets of particles.  The 2-dimensional projected
mass profile, $M_{\sigma_v}(R)$, is shown in Fig.\ 10 and the 3-dimensional
mass profile, $M_{\sigma_v}(r)$, is shown in Fig.\ 11.  As in Figs.\
8 and 9, circles refer to values of $\sigma_v$ computed using the large
cumulative annuli and squares refer to values of $\sigma_v$ computed
using the thin, independent annuli.  Both the projected
mass profiles and the 3-dimensional mass profiles estimated directly from the 
particle velocity dispersions show scale-dependent deviations from the
true mass profile. In this case the cluster mass is overestimated at 
very  small
radii and underestimated over most of the cluster.

\vspace*{-1.0truecm}
\hbox{~}
\centerline{\psfig{file=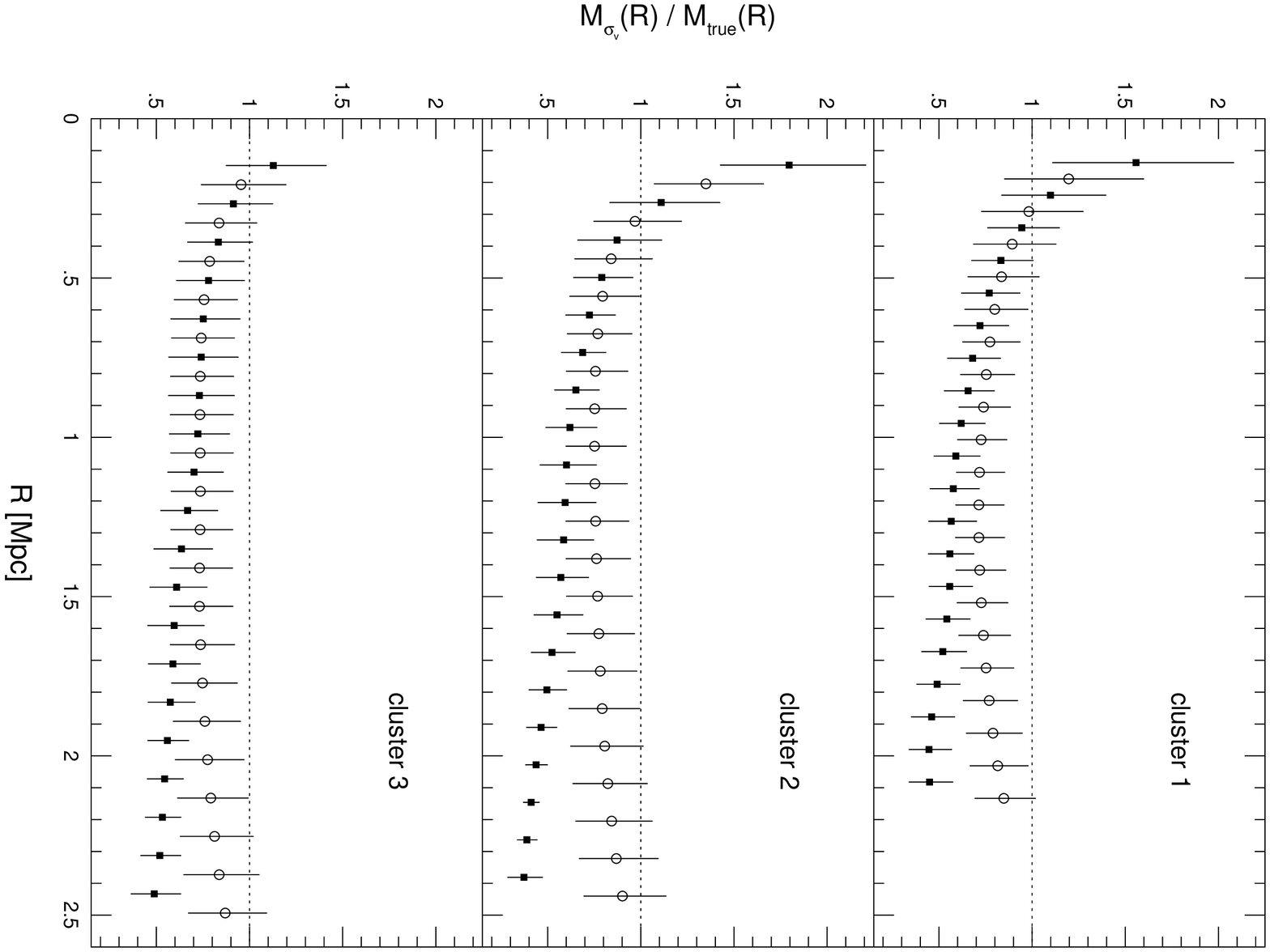,angle=270,width=4.5in}}
\vspace*{1.7truecm}
\noindent{\scriptsize
\addtolength{\baselineskip}{-3pt}
Fig.~10:
The 2-dimensional, projected cluster mass profile obtained
directly from
the measured particle velocity dispersion (assuming the clusters to
be isothermal spheres) compared to the true cluster mass profile.
Solid squares indicate that the value of $\sigma_v$ used in equation (8)
was determined from independent annuli of differential radius $R$;
open circles indicate that the value of $\sigma_v$ was determined
from large, cumulative annuli with outer radii of $R_{\rm max} = R$.
Error bars show the 1-$\sigma$ dispersion in $M(R)$ amongst the 1000 projections
from which the mean line-of-sight velocity dispersion was computed.
\label{fig10}
\addtolength{\baselineskip}{3pt}
}

\bigskip
Lastly, shown in Fig.\ 12 is a 3-dimensional mass profile computed for
the clusters using a virial mass estimator.  The classical cluster
virial mass estimator is:
\begin{equation}
M = \frac{3\pi \sigma_v^2 R_e}{2G}
\end{equation} 
where $R_e$ is the mean effective radius as projected on the sky:
\begin{equation}
R_e^{-1} \equiv \frac{1}{N^2} \sum_{i < j}^{N} \frac{1}
{| \vec{R}_i - \vec{R}_j |}
\end{equation}
and $N$ is the number of galaxies in the cluster.  Again, we
cannot resolve the individual dark matter halos of member galaxies and,
so, the virial analysis was performed on the clusters using random subsets of
the particles.  Particles contained within
concentric spheres of radius $r$ centered on the cluster centers of
mass were viewed from 1000 random orientations and $\sigma_v$
and $R_e^{-1}$ were computed for each orientation.  Values 
of $r$ were increased
incrementally to $r_{\rm max} = r_{200}$, and $M(r)$, the mass contained within
the concentric spheres, was
computed using equation (13) above.  From Fig.\ 12,
the virial mass estimator leads to a scale-dependent deviation from
the true 3-dimensional mass profile in the sense that
the cluster mass is overestimated
on small scales.  On large scales (and, in particular, near the
``edges'' of the clusters), however, the virial mass estimator reproduces the 
true cluster mass quite well.

\vspace*{-1.0truecm}
\hbox{~}
\centerline{\psfig{file=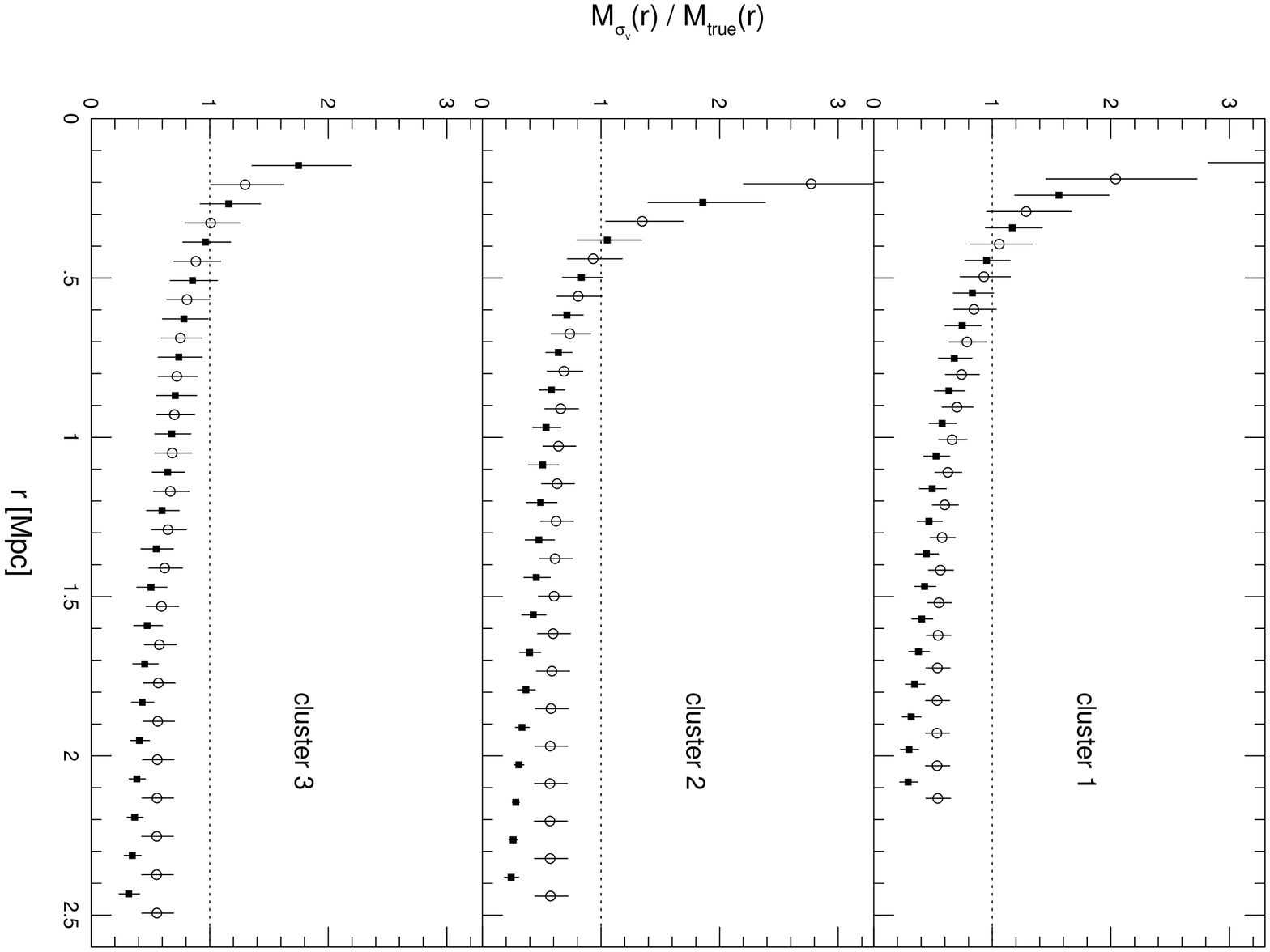,angle=270,width=4.5in}}
\vspace*{1.7truecm}
\noindent{\scriptsize
\addtolength{\baselineskip}{-3pt}
Fig.~11:
The 3-dimensional cluster mass profile obtained directly from
the measured particle velocity dispersion (assuming the clusters to
be isothermal spheres) compared to the true cluster mass profile.
Solid squares indicate that the value of $\sigma_v$ used in equation (7)
was determined from independent annuli of differential radius $R$;
open circles indicate that the value of $\sigma_v$ was determined
from large, cumulative annuli with outer radii of $R_{\rm max} = R$.
Error bars show the 1-$\sigma$ dispersion in $M(r)$ amongst the 1000 projections
from which the mean line-of-sight velocity dispersion was computed.
\label{fig11}
\addtolength{\baselineskip}{3pt}
}

\bigskip
It should be noted that gravitational 
force softening in the numerical simulation
will, necessarily, affect dynamical mass estimates of simulated
objects (see, e.g.,
Tormen, Bouchet \& White 1997).  That is, on scales smaller
than or of order the smoothing length, the mass will be severely overestimated
simply due to numerical effects.  We have, therefore, restricted our analyses
to radii at which the effects of force softening on the mass
estimate should be small.  In particular, any overestimate of
the mass caused by numerical effects is expected to be at most of order
3\% to 4\% in the innermost radial bins and will drop rapidly to
zero for the bins with larger radii.

Cen (1997) and Reblinsky \& Bartelmann (1999) have also investigated the
virial masses obtained for numerical clusters, though not for objects
as massive as those presented here.  Reblinsky \& Bartelmann (1999)
find that the virial mass
severely overestimates the true masses of clusters whose masses are less than
a few times $10^{14} M_\odot$.  The degree of overestimation decreases
with increasing cluster mass, however, and appears to converge in the mean
to the true cluster mass for their most massive objects.  Reblinsky \&
Bartelmann's results are broadly consistent with those of
Cen (1997), though differences in the procedures used to select
and analyse the clusters makes direct comparisons between the two
not entirely straightfoward.  Direct comparisons between our results
and those of Cen (1997) and Reblinsky \& Bartelmann (1999) are also
not straightforward.

\vspace*{-1.0truecm}
\hbox{~}
\centerline{\psfig{file=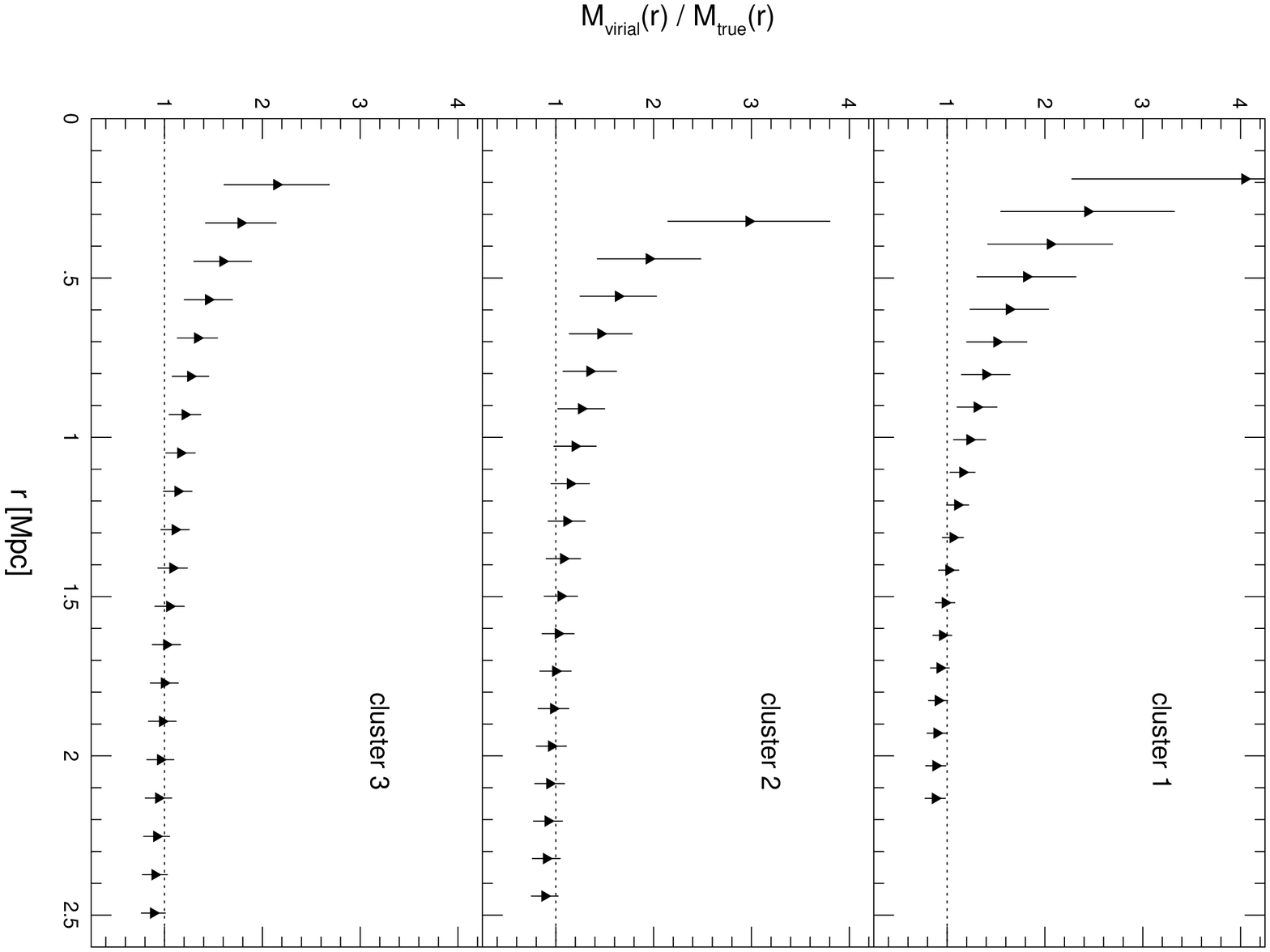,angle=270,width=4.5in}}
\vspace*{1.7truecm}
\noindent{\scriptsize
\addtolength{\baselineskip}{-3pt}
Fig.~12:
The 3-dimensional cluster mass profile obtained from
the classical virial mass estimator compared to the true cluster
mass profile.  Particles contained within concentric spheres of
radius $r$ centered on the cluster center of mass were used to
determine the mean values of $R_e$ and $\sigma_v$ required for the
evaluation of equation (13).
Error bars show the 1-$\sigma$  dispersion in $M(r)$ amongst 1000
random projections.
\label{fig12}
\addtolength{\baselineskip}{3pt}
}

\bigskip
In our analyses above we have expressly calculated $R_e$ for each of
the subsets
of the particles, whereas Cen (1997) and Reblinksy \& Bartelmann (1999)
do not.  Also, we have selected our clusters from a 3-dimensional mass
distribution while
Cen (1997) and Reblinsky \& Bartelmann (1999) select their clusters
based on 2-dimensional projections and the assignment of luminous galaxies
to a random subset of the particles in their simulations.  As such, their
analyses attempt to at least partially address the issue of contamination
by interloper galaxies and false detections of clusters in the limit of
realistic observational data.   In contrast, our results above are effectively
derived in the limit of ideal data (i.e. the values of $R_e$ and $\sigma_v^2$
are computed from objects which are known a priori to be contained within
the cluster under investigation).

\section{Discussion}

The cluster mass results which are the most relevant for direct comparison to 
observational investigations are those that were obtained using
large, cumulative annuli (i.e.\ the shear and particle velocity 
dispersion averaged over large scales in the cluster) as well as the virial
estimate in which $R_e$ is the effective radius determined for the
``entire'' cluster.  Although in principle the shear and velocity dispersion
can be measured at independent radii in observed clusters, the data
are generally too sparse and noisy for this to be practicable. 
(See, however, Bonnet et al.\ (1994), Tyson \& Fischer (1995),
Squires et al.\ (1996b), Fisher \& Tyson (1997) and Carlberg et al.\
(1997b) for exceptions to this.)

The mass profiles plotted in Figs.\ 8 through 12 extend to a maximum
cluster radius
equal to $r_{200}$ (or $R_{200}$ in the case of the projected mass
profiles).  In all cases $M(r)$ and $M(R)$ in these figures
refer to the mass contained within a 3-dimensional radius, $r$, or a
projected radius, $R$.
Since we have defined the clusters to consist of all 
particles inside of a radius $r_{200}$, we will define the total mass of
a cluster to be the mass contained within this radius, $M(r_{200})$.  The 
total mass obtained for each  cluster from each of the
estimators is, therefore, indicated by the points in Figs.\ 8 through 12
that are plotted at the largest occurring values of the radius.  

In terms of estimating the total cluster mass (i.e.\ the mass of
the cluster contained within a 3-dimensional radius of
$r_{200}$), the classical virial estimator is found to
be very successful.  The total mass of the cluster is systematically
underestimated, but only by $\sim 10$\%.  This result is somewhat 
surprizing given the fact that within $r_{200}$ the cluster mass distributions
are not perfectly smooth and substructure exists at a significant
level.  Additionally, moment of inertia analyses performed using all
particles within a radius $r_{200}$ of the cluster centers of mass show
the cluster mass distributions to be clearly triaxial, rather than spherical
(see Brainerd, Goldberg \& Villumsen 1998 for the relevant discussions).
Our result, therefore, suggests that at least in the limit of ideal
data the classical virial mass estimator is
quite robust to modest deviations from pure spherical symmetry and the
presence of substructure within a cluster.

The ``isothermal'' dynamical mass estimate, 
in which the measured 
line of sight velocity dispersion is used to infer the mass under the
assumption of an isothermal potential, yields a poor estimate of
the total mass of the cluster.  The value of 
$M(r_{200})$ is underestimated by $\sim 40$\% for the case in which $\sigma_v$ 
is determined
from an average over the entire cluster and is underestimated by 
$\sim 70$\% for the case in which  $\sigma_v$ is computed at a projected
radius of $R = R_{200}$ (i.e.\ Fig.\ 11).

Provided the mean shear
used to infer the cluster velocity dispersion is computed using a large,
cumulative annulus in which the shear is averaged over the entire cluster,
the weak lensing estimate of $M(r_{200})$ is found to be in
excellent agreement with the total cluster mass
(i.e.\ Fig.\ 9, open circles).  In contrast, however,
the shear measured solely at a radius of
$R=R_{200}$ yields a $\sim 25$\%
overestimate of the total cluster mass (i.e.\  Fig.\ 9, solid squares)
due to the fact that the clusters are finite in extent, rather than infinite.

Because of its promise to yield direct measurements of the masses
of galaxy clusters independent of dynamics and hydrodynamics, weak
lensing mass estimates of cluster masses are currently of particular 
interest.  
Given the fact that most high-quality observations of the
weak lensing shear due to clusters have been obtained only on 
relatively small scales
(i.e.\ radii significantly less than 1 or 2 Mpc), 
Figs.\ 8 and 9 suggest some caution
regarding the interpretation of recent cluster mass estimates that are
based on a measurement
of an average value of the weak shear together with
an assumption of an isothermal
cluster potential.  In particular, a measurement of the mean shear
in which the mean is computed within an aperture whose outer radius
is significantly less than $R_{200}$ yields a mass estimate that 
differs systematically from the true mass.  For example,
the projected mass within a radius of 0.5~Mpc, $M(R=0.5 \; {\rm Mpc})$, 
is underestimated by $\sim 40$\%
and the 3-dimensional contained mass, $M(r=0.5 \; {\rm Mpc})$, is 
underestimated by $\sim 35$\%.  Interestingly, in an
analysis of observed cluster lensing data, Wu et al.\ (1998) found that
weak lensing mass estimates that were performed over small cluster radii 
did seem to underestimate the contained mass in a systematic
manner.

The results for $M_{\rm lens}(r)$ shown in Fig.\ 9 are, however,
encouraging
at large cluster radii.  In particular, with the advent of large 
format CCD cameras capable of wide-field imaging, it will be possible
to measure the weak shear due to lensing clusters at radii of order
a few Mpc in a reasonably routine fashion.   An example of such deep
wide-field imaging is the
data obtained with the UH 8K CCD mosaic camera which has recently resulted
in a detection of large scale coherent weak shear in the images
of $\sim 30,000$ faint background galaxies due to lensing by the
supercluster MS0302+17 (Kaiser et al.\ 1998).
Given the apparent
universality of the Navarro, Frenk \& White density profile (i.e.\
dissipationless collapse generically leads to the formation of
an object with an NFW profile), our 
results suggest that it will be possible
to estimate a total
3-dimensional cluster mass fairly accurately with wide-field
imaging simply by computing the mean of the shear over the entire
cluster and adopting an isothermal lens potential.  In the short
term, such observations
will hopefully provide a resolution to the remaining discrepancies 
between cluster masses estimated from weak lensing and virial techniques.
(This is, of course, providing that the redshift distribution of the 
lensed galaxies is well-constrained and is not, in itself, a large
source of uncertainty in the interpretation of the observed shear.) 
In the long term, large surveys from which the weak lensing shear can be
detected out to large cluster radii should have the ability 
to yield uniform samples
of objects, including a reasonably accurate mass-selection criterion,
without necessarily requiring a full reconstruction of
the density profile of each
individual lensing cluster.  

\section*{Acknowledgments}

A generous allocation of computing resources on Boston University's
Origin2000, support under NSF contract
AST-9616968 (TGB and COW) and NSF Graduate Fellowships (DMG and
COW) are gratefully acknowledged.


\begin{references}

\reference{bfc97} Bahcall, N. A., Fan, X., \& Cen, R. 1997, ApJ, 485, L53

\reference{bc93} Bahcall, N. A. \& Cen, R. 1993, \apjl, 407, L49

\reference{bld} Bahcall, N. A., Lubin, L. M. \& Dorman, V. 1995, \apj, 447, L81

\reference{msb95} Bartelmann, M. 1995, A\&A, 303, 643

\reference{msb96} Bartelmann, M., Narayan, R., Seitz, S. \& Schneider,
P. 1996, \apj, 464, L115

\reference{bt87} Binney, J. \& Tremaine, S. 1987, Galactic
Dynamics (Princeton: Princeton University Press)

\reference{bsbv} Blandford, R. D., Saust, A. B., Brainerd, T. G., \&
Villumsen, J. V. 1991, \mnras, 251, 600

\reference{bon94} Bonnet, H., Mellier, Y. \& Fort, B. 1994, \apj, 427, L83

\reference{bs97} Bower, R. G. \& Smail, I. 1997, \mnras, 290, 292

\reference{paper1} Brainerd, T. G., Goldberg, D. M. \& Villumsen, J. V.
1998, \apj, 502, 505

\reference{t6} Bromley, B. C., Warren, M. S., Zurek, W. H., \& Quinn, P. J.
1995, AIP Conference Proceedings 336, 433

\reference{bunn} Bunn, E. \& White, M. 1997, \apj, 480, 6

\reference{ray2} Carlberg, R. G., Yee, H. K. C. \& Ellingson, E. 1994,
\apj, 437, 63

\reference{ray1} Carlberg, R. G., Yee, H. K. C. \& Ellingson, E. 1997,
\apj, 478, 462

\reference{cnoc96} Carlberg, R. G., Yee, H. K. C., Ellingson, E., Abraham,
R., Gravel, P., Morris, S., \& Pritchet, C. J. 1996, \apj, 462, 32

\reference{cnoc97a} Carlberg, R. G., Yee, H. K. C., Ellingson, E., 
Morris, S. L., Abraham, R., Gravel, P., Pritchet, C. J., Smecker-Hane, T.,
Hartwick, F. D. A., Hesser, J. E., Hutchings, J. B., \& Oke, J. B. 1997a,
\apj, 476, L7

\reference{cnoc97b} Carlberg, R. G., Yee, H. K. C., Ellingson, E., 
Morris, S. L., Abraham, R., Gravel, P., Pritchet, C. J., Smecker-Hane, T.,
Hartwick, F. D. A., Hesser, J. E., Hutchings, J. B., \& Oke, J. B. 1997b,
\apj, 485, L13

\reference{cen97} Cen, R. 1997, \apj, 485, 39

\reference{cl96} Cole, S. M. \& Lacey, C. G. 1996, MNRAS, 281, 716

\reference{dml} Dahle, H., Maddox, S. J. \& Lilje, P. B. 1994, \apj, 435, L79

\reference{djf} David, L. P., Jones, C. \& Forman, W. 1995, \apj, 445, 578

\reference{dc91} Dubinski, J. \& Carlberg, R. 1991, ApJ, 378, 496

\reference{eke96} Eke, V. R., Cole, S. \& Frenk, C. S. 1996, \mnras, 282, 263

\reference{fah94} Fahlman, G., Kaiser, N., Squires, G. \& Woods, D. 1994,
\apj, 289, L1

\reference{fbc97} Fan, X., Bahcall, N. \& Cen, R. 1997, ApJ, 415, L17

\reference{fischer97} Fischer, P., Bernstein, G., Rhee, G., \& Tyson, J. A.
1997, \apj, 113, 521

\reference{ft97} Fischer, P. \& Tyson, J. A. 1997, \aj, 114, 14

\reference{ben98} Ghigna, S., Moore, B., Governato, F., Lake, G.,
Quinn, T., \& Stadel, J. 1998, \mnras, 300, 146

\reference{gt} Gott, J. R. \& Turner, E. L. 1976, \apj, 209, 1

\reference{gunn} Gunn, J. E. 1978, in Observational Cosmology, ed.
A. Maeder, L. Martinet, \& G. Tammann (Sauverny: Geneva Obs.), 1

\reference{nk95} Kaiser, N. 1995, \apj, 439, L1

\reference{ks93} Kaiser, N. \& Squires, G. 1993, \apj, 404, 441

\reference{ksb95} Kaiser, N., Squires, G. \& Broadhurst, T. 1995, \apj, 449,
460

\reference{jpk96} Kneib, J.-P., Ellis, R. S., Smail, I., Couch, W. J.,
\& Sharples, R. M. 1996, \apj, 471, 643

\reference{uh8k} Kaiser, N., Wilson, G., Luppino, G., Kofman, L.,
Gioia, I., Metzger, M., \& Dahle, H. 1998, ApJ submitted
(astro-ph/9809268)

\reference{lb93} Lubin, L. M. \& Bahcall, N. A. 1993, ApJ, 415, L17

\reference{lp97} Luppino, G. \& Kaiser, N. 1997, \apj, 475, 20

\reference{m94} Mellier, Y., Dantel-Fort, M., Fort, B., \& Bonnet, H.
1994, A\&A, 289, L15

\reference{jordi} Miralda-Escud\'e, J. \& Babul, A. 1995, \apj, 449, 18

\reference{nfw95} Navarro, J. F., Frenk, C. S. \& White, S. D. M. 1995, 
\mnras, 275, 720

\reference{nfw96} Navarro, J. F., Frenk, C. S. \& White, S. D. M. 1996, 
\apj, 462, 563

\reference{nfw97} Navarro, J. F., Frenk, C. S. \& White, S. D. M. 1997, 
\apj, 490, 493

\reference{owb99} Oaxaca Wright, C. \& Brainerd, T. G. 1999, in preparation

\reference{rgh} Ramella, M., Geller, M. J. \& Huchra, J. P. 1989,
\apj, 344, 57

\reference{rb99} Reblinsky, K. \& Bartelmann, M. 1999, A\&A, in press
(astro-ph/9902153)

\reference{ps95} Schneider, P. 1995, A\&A, 302, 639

\reference{dasB} Schneider, P., Ehlers, J., \& Falco, E. E. 1992,
Gravitational Lensing (Berlin: Springer-Verlag)

\reference{ss95a} Schneider, P. \& Seitz, C. 1995, A\&A, 294, 411

\reference{schw} Schwarzschild, M. 1954, \aj, 59, 273

\reference{ss95b} Seitz, C. \& Schneider, P. 1995, A\&A, 297, 287

\reference{s1996} Seitz, C., Kneib, J.-P., Schneider, P., \&
Seitz, S. 1996, A\&A, 314, 707

\reference{ss96} Seitz, S. \& Schneider, P. 1996, A\&A, 305, 383

\reference{ssb98} Seitz, S., Schneider, P. \& Bartelmann, M. 1998,
\apj, 337, 325

\reference{sd95} Smail, I. \& Dickinson, M. 1995, \apj, 455, L99

\reference{irs94} Smail, I., Ellis, R. S., Fitchett, M. J. 1994, \mnras,
270, 245

\reference{irs95} Smail, I., Ellis, R. S., Fitchett, M. J., \&
Edge, A. C. 1995, \mnras, 273, 277

\reference{irs97} Smail, I., Ellis, R. S., Dressler, A., Couch, W., Oemler,
A., Sharples, R. M., \& Butcher, H. 1997, ApJ, 479, 70

\reference{smith} Smith, S. 1936, \apj, 83, 29

\reference{sk96} Squires, G. \& Kaiser, N. 1996, \apj, 473, 65

\reference{s1996a} Squires, G., Kaiser, N., Babul, A., Fahlman, G., Woods,
D., Neumann, D. M., \& B\"ohringer, H. 1996a, \apj, 461, 572

\reference{s1996b} Squires, G., Kaiser, N., Fahlman, G., Babul, A.,
\& Woods, D. 1996b, \apj, 469, 73

\reference{tbw97} Tormen, G., Bouchet, F. \& White, S. D. M. 1997, MNRAS,
286, 865

\reference{twv90} Tyson, J. A., Wenk, R. A. \& Valdes, F. 1990, \apj,
349, L1

\reference{tf95} Tyson, J. A. \& Fischer, P. 1995, \apj, 466, L55

\reference{eli} Waxman, E. \& Miralda-Escud\'e, J. 1995, \apj, 451, 451

\reference{white93} White, S. D. M., Efstathiou, G. \& Frenk, C. S. 1993, 
\mnras, 262, 1023

\reference{viana96} Viana, T. P. \& Liddle, A. R. 1996, \mnras, 281, 323

\reference{jvv} Villumsen, J. V. 1989, \apjs, 71, 407

\reference{wu98} Wu, X.-P., Chiueh, T., Fang, L.-Z., \& Yan-Jie, X. 1998,
\mnras, in press

\reference{fritz1} Zwicky, F. 1933, Helv. Phys. Acta, 6, 110

\reference{fritz2} Zwicky, F. 1937, \apj, 86, 217

\end{references}
\end{document}